\documentclass[aps,pra,preprint,groupedaddress]{revtex4-1}
\usepackage{amsmath}
\usepackage{graphicx}

\begin{document}


\title{Are vector vortex beams endowed with any entanglement?}


\author{Chun-Fang Li}
\email[]{cfli@shu.edu.cn}
\affiliation{Department of Physics, Shanghai University, 99 Shangda Road, 200444 Shanghai, China}


\date{\today}

\begin{abstract}

Polarization of light beams is one of the most important physical phenomena. But up till now it was only described in the paraxial approximation in which it is considered to be a single degree of freedom that is characterized by the local Stokes parameters over the transverse plane. Based on such a description, vector vortex beams are considered to be entangled in polarization and spatial mode.
Here we show that there is not any entanglement in a large class of representative vector vortex beams, including the well-known cylindrical-vector beams.
This is achieved by developing an approach to exactly characterize the polarization of a general beam.
It is found that the Stokes parameters, when generalized rigorously to a general beam in momentum space, are physical quantities with respect to a natural coordinate system.
The so-called Stratton vector determining the natural coordinate system fixes a natural representation for the polarization in which the Pauli matrices represent the intrinsic degree of freedom of the polarization with respect to the natural coordinate system. As a result, the Stratton vector itself shows up as another degree of freedom of the polarization.
From this point of view, the light beams specified by a Stratton vector parallel to the propagation axis as well as by the eigenvalues of the Pauli matrix $\hat{\sigma}_1$ are precisely vector vortex beams. They are not endowed with any entanglement.

\end{abstract}


\maketitle


\newpage
\section{Introduction}

Vector vortex beams refer to light beams the direction of whose electric or magnetic field varies over the transverse profile. Of these, cylindrical-vector beams \cite{Youn-B, Zhan} have rotationally invariant electric or magnetic fields. Such a beam is considered to be non-separable or entangled in polarization and spatial mode, classically \cite{Topp-AMGL, Aiel-TMGL, Mili-NLNA, McLa-KF} or quantum-mechanically \cite{D'Am-CG}. For example, a cylindrical-vector beam is viewed as a superposition of two different, orthogonally polarized spatial modes \cite{Zhan, Topp-AMGL, Aiel-TMGL, Mili-NLNA, McLa-KF, D'Am-CG, Stal-S, Fick-LRZ}.
The problem is that this conclusion is drawn on the basis of a paraxially approximate description of the vector vortex beams in which the axial component of the electric or magnetic field \cite{Youn-B} is ignored. The notion of polarization in such a description is postulated to be a single degree of freedom \cite{Fick-LRZ, Kwia-MWZ, Mich-WZ, Kies-SWUW, Simo-SG, Pan-CL, Bayr-SCB, Ecker}.
The local Stokes parameters \cite{Meji-MPM, Beck-BA, Card} in the transverse plane are taken as the physical observables \cite{Coll, Gold} to characterize the polarization state of paraxial beams.

First of all, what bears the name of polarization in a light beam is the direction of the intact electric or magnetic field \cite{Born-W} rather than the direction of their approximations.
Secondly, the Stokes parameters can only be strictly defined for the polarization of a plane wave \cite{Jauc-R}. Different from a vector vortex beam, a plane wave has no axial component.
Even more noteworthy is that the Pauli matrices that express \cite{Fano} the Stokes parameters of a plane wave do not act on the electric field. Instead, they act in the Hilbert space of two-element Jones vectors \cite{Jones}.
Therefore, the Jones vector of a plane wave should be distinguished from its electric field though the two elements of the Jones vector result from the projection of the electric field onto the polarization bases.
According to Jauch and Rohrlich \cite{Jauc-R}, the Jones vector of a plane wave is some kind of spinor. Hereafter we will call it the Jones spinor. Unfortunately, this spinor has not received the attention it deserves. Its physical meaning remains unclear. Consequently, the Stokes parameters it gives through the Pauli matrices are simply depicted on the surface of the Poincar\'{e} sphere \cite{Coll, Gold, Born-W, Jauc-R, Fano}.
This naturally raises the question of whether the paraxially approximate description of vector vortex beams can properly convey their polarization states.

To address this issue, one has to know how to exactly characterize the polarization state of a general beam.
As mentioned above, the Stokes parameters can only be strictly defined for plane waves.
If they were able to exactly characterize the polarization of a plane wave as is commonly assumed \cite{Gold, Cald, Berr-GL}, it would be straightforward to get an exact characterization of the polarization of a general beam by generalizing the Stokes parameters in momentum space on the basis of the Fourier transformation. The fact that no such approach exists \cite{Meji-MPM} in the literature implies that there should be something wrong with that assumption.
We will show in this paper that the root of the problem really lies with the peculiarity of the above-mentioned Jones spinor.

It is revealed that the generalized Stokes parameters are quantities with respect to some natural coordinate system (NCS) rather than with respect to the laboratory coordinate system (LCS). As a result, the Stokes parameters at each point in momentum space depend on the choice of the NCS.
The situation encountered here is similar to that encountered in Einstein's special theory of relativity \cite{Gold1} in which the physical quantities of a particle such as the momentum and energy depend on the choice of inertial reference frame.
By this it is meant that the momentum-space Jones spinor that gives the Stokes parameters through the Pauli matrices is defined over the NCS.
In order to completely determine the Jones spinor and hence the Stokes parameters, one needs to figure out a way to specify the NCS at every momentum.
To be honest, this is not the only case that requires specifying such a NCS. One also needs to do so in representing a light beam with a Fourier integral. Stratton \cite{Stra} seemed to be first to deal with this problem by introducing a constant real unit vector. The same technique was later used independently by others \cite{Gree-W, Patt-A, Davi-P}.
It is expounded that so introduced unit vector, called the Stratton vector (SV), just reflects what is needed to exactly characterize the polarization of a general beam.
Contrary to the above-mentioned postulation, the polarization actually involves two different kinds of degrees of freedom.
One is represented by the Pauli matrices with respect to the NCS. The other is the SV that plays the role of specifying the NCS.
When viewed in terms of these two degrees of freedom, the polarization of the vector vortex beams is not entangled with any other degrees of freedom.

\section{Stokes parameters in momentum space: Introduction of SV}

Without loss of generality, we consider a general monochromatic beam propagating along the $z$ axis in free space. Its complex-valued electric field, when expressed as
$\mathbf{E}(\mathbf{x}) \exp(-i \omega t)$,
can be expanded in terms of the plane-wave modes in the following way,
\begin{equation}\label{FT}
  \mathbf{E} (\mathbf{x})
 =\frac{1}{2 \pi} \int_0^k k_\rho d k_\rho \int_0^{2 \pi} d \varphi
  \mathbf{e}(\mathbf{k}) \exp(i \mathbf{k} \cdot \mathbf{x}),
\end{equation}
where $\omega=c k$, $c$ is the speed of light in free space, $k=|\mathbf{k}|$,
$\mathbf{e} (\mathbf{k})$ is the electric field in momentum space, $k_\rho$ is the radial component of the wavevector $\mathbf k$ in cylindrical coordinates, and $\varphi$ is the azimuthal angle.
As is well known, the electric field $\mathbf e$ in momentum space satisfies
\begin{equation}\label{TC-MS}
    \mathbf{k} \cdot \mathbf{e}(\mathbf{k})=0,
\end{equation}
by virtue of the Maxwell equation
\begin{equation}\label{ME}
\nabla \cdot \mathbf{E} =0.
\end{equation}
To separate out the polarization from the intensity \cite{Coll, Gold, Born-W, Jauc-R, Fano}, we split the strength factor off from the momentum-space electric field by writing
\begin{equation}\label{vec-e}
    \mathbf{e}(\mathbf{k})=e(\mathbf{k}) \mathbf{a}(\mathbf{k}),
\end{equation}
where the strength factor $e(\mathbf{k})$ satisfies
$|e(\mathbf{k})|=|\mathbf{e}(\mathbf{k})|$.
The separated unit-vector function $\mathbf{a} (\mathbf{k})$, known as the polarization vector \cite{Akhi-B, Cohe-DG}, represents the state of polarization.
What is noteworthy is that it is not independent of the momentum. It is constrained by the transversality condition
\begin{equation}\label{TC}
\mathbf{k} \cdot \mathbf{a} (\mathbf{k})=0,
\end{equation}
in accordance with Eq. (\ref{TC-MS}).
The polarization vector $\mathbf{a}$ in momentum space should not be confused with the electric-field vector (\ref{FT}) in position space.
It is of course true that the vectorial property of $\mathbf a$ determines the vectorial property of $\mathbf{E}$ in the sense that if $\mathbf{a}$ were not a vectorial function, $\mathbf{E}$ would not be a vectorial function, either.
Nevertheless, the distribution of $\mathbf E$ in position space depends not only on the polarization vector $\mathbf{a}$ but also on the strength factor $e$.

Equation (\ref{TC}) means that at each point $\mathbf k$ in momentum space there exists such a pair of mutually-perpendicular unit vectors, $\mathbf u$ and $\mathbf v$, that form, with $\mathbf k$, a NCS, satisfying
\begin{equation}\label{triad}
    \mathbf{u} \cdot  \mathbf{v}=0,                   \quad
    \mathbf{u} \times \mathbf{v}= \frac{\mathbf k}{k}.
\end{equation}
One can choose them as the polarization bases to expand the polarization vector $\mathbf{a}$ at that momentum as
$
\mathbf{a}=\alpha_1 \mathbf{u} +\alpha_2 \mathbf{v}.
$
The two expansion coefficients make up the Jones spinor
$
\alpha=\Big(\begin{array}{c}
              \alpha_1 \\
              \alpha_2
            \end{array}
       \Big)
$
at the associated momentum, which is normalized as
$\alpha^\dag \alpha=1$.
Introducing the matrix
\begin{equation}\label{varpi}
   \varpi \equiv (
                  \begin{array}{cc}
                     \mathbf{u} & \mathbf{v} \\
                  \end{array}
                 ),
\end{equation}
which contains the polarization bases as the column vectors, one can express the polarization vector simply in terms of the Jones spinor as
\begin{equation}\label{QUT1}
\mathbf{a}=\varpi \alpha.
\end{equation}
Matrix (\ref{varpi}) has the property
\begin{equation}\label{QU1}
    \varpi^\dag \varpi=I_2
\end{equation}
by virtue of Eqs. (\ref{triad}), where the superscript $\dag$ stands for the complex transpose and $I_2$ is the 2-by-2 unit matrix.
Multiplying Eq. (\ref{QUT1}) by $\varpi^\dag$ on the left and making use of Eq. (\ref{QU1}), one is able to write the Jones spinor in terms of the polarization vector as
\begin{equation}\label{QUT2}
    \alpha =\varpi^\dag \mathbf{a}.
\end{equation}
With the Jones spinor (\ref{QUT2}), it is straightforward to follow the procedure that is used for plane waves to define the Stokes parameters for the polarization state $\mathbf{a}$ as follows,
\begin{equation}\label{SP}
    s_i=\alpha^\dag \hat{\sigma}_i \alpha, \quad i=1,2,3,
\end{equation}
where $\hat{\sigma}_i$ are the Pauli matrices \cite{Coll, Fano}
\begin{equation}\label{PM}
\hat{\sigma}_1=\bigg(\begin{array}{cc}
1 &  0 \\
0 & -1
\end{array}
\bigg),                 \quad
\hat{\sigma}_2=\bigg(\begin{array}{cc}
0 & 1 \\
1 & 0
\end{array}
\bigg),                 \quad
\hat{\sigma}_3=\bigg(\begin{array}{cc}
0 & -i \\
i &  0
\end{array}
\bigg).
\end{equation}
Generally speaking, the Jones spinor (\ref{QUT2}) and hence the Stokes parameters (\ref{SP}) are functions of the momentum.

However, the Stokes parameters so far have not yet been completely determined. As is known, for a monochromatic plane wave propagating along the $z$ axis, it is always possible to choose the unit vectors along the $x$ and $y$ axes of the LCS as the polarization bases.
But for the general beam (\ref{FT}), the polarization bases $\mathbf u$ and $\mathbf v$ at different momenta can not be the same. Particularly noteworthy is the fact that they are arbitrary to the extent that a rotation about the associated momentum $\mathbf k$ can be performed.
To completely determine the momentum-space Stokes parameters through Eqs. (\ref{SP}) and (\ref{QUT2}), one has to figure out a way to determine the polarization bases, or the transverse axes of the NCS $\mathbf{uvw}$, at different momenta.
Fortunately, it was first shown by Stratton \cite{Stra} and later by others \cite{Gree-W, Patt-A, Davi-P} that this may be done consistently by introducing a constant real unit vector
$\mathbf I$ in the following way,
\begin{equation}\label{TA}
    \mathbf{u} =\mathbf{v} \times \frac{\mathbf k}{k},                \quad
    \mathbf{v} =\frac{\mathbf{k} \times \mathbf{I}}{|\mathbf{k} \times \mathbf{I}|}.
\end{equation}
So introduced unit vector is what we call the SV. Indeed, it is easy to check that the unit vectors $\mathbf u$ and $\mathbf v$ determined through these two equations by any SV satisfy Eq. (\ref{triad}).
One might object to such a seemingly artificial definition, especially after noticing that so defined polarization bases are indeterminate at the momentum that is parallel to $\mathbf I$.
Nevertheless, as we will see in Section \ref{no-entangle}, it is this indeterminacy that underlies the so-called ``polarization singularity'' \cite{D'Am-CG, Fick-LRZ, Bomz-BK} in cylindrical-vector beams.

From Eqs. (\ref{TA}) it follows that the Jones spinor of a polarization state given by Eqs. (\ref{QUT2}) and (\ref{varpi}) is always associated with some SV.
As a result, the Stokes parameters (\ref{SP}) it gives through the Pauli matrices (\ref{PM}) are also associated with that SV.
We are thus faced with the problems as to whether and how the SV-associated Stokes parameters can characterize the polarization state of the general beam (\ref{FT}).

\section{Polarization wavefunction in natural representation}

\subsection{Transformation of Jones spinor under change of SV}

To deal with these problems, let us first look at how the Jones spinor of a polarization state depends on the choice of the SV.
Consider a different SV, say $\mathbf{I}'$. In this case, the transverse axes at momentum $\mathbf k$ take the form
\[
    \mathbf{u}' =\mathbf{v}' \times \frac{\mathbf k}{k}, \hspace{5pt}
    \mathbf{v}' =\frac{\mathbf{k} \times \mathbf{I}'}{|\mathbf{k} \times \mathbf{I}'|}.
\]
According to Eq. (\ref{QUT2}), the Jones spinor of the same polarization state $\mathbf a$ in association with the primed SV reads
\begin{equation}\label{JV'}
    \alpha'=\varpi'^{\dag} \mathbf{a},
\end{equation}
where
$\varpi'=(\begin{array}{cc}
            \mathbf{u}' & \mathbf{v}'
          \end{array}
         ).
$
As mentioned before, the primed transverse axes
$\mathbf{u}'$ and $\mathbf{v}'$
are related to the unprimed ones $\mathbf{u}$ and $\mathbf{v}$ by a rotation about the associated momentum $\mathbf k$.
Denoted by $\Phi$, the rotation angle obeys
\begin{subequations}\label{R-TA}
\begin{align}
  \mathbf{u}' & = \mathbf{u} \cos \Phi +\mathbf{v} \sin \Phi,   \\
  \mathbf{v}' & =-\mathbf{u} \sin \Phi +\mathbf{v} \cos \Phi.
\end{align}
\end{subequations}
These two equations can be combined into one single equation of the following form,
\begin{equation}\label{R-LTA1}
    \varpi'=\exp [-i(\hat{\mathbf \Sigma} \cdot \mathbf{w}) \Phi] \varpi,
\end{equation}
where
\begin{equation*}
    \hat{\Sigma}_x=\Bigg(\begin{array}{ccc}
                           0 & 0 &  0 \\
                           0 & 0 & -i \\
                           0 & i &  0
                         \end{array}
                   \Bigg),              \quad
    \hat{\Sigma}_y=\Bigg(\begin{array}{ccc}
                            0 & 0 & i \\
                            0 & 0 & 0 \\
                           -i & 0 & 0
                         \end{array}
                   \Bigg),              \quad
    \hat{\Sigma}_z=\Bigg(\begin{array}{ccc}
                           0 & -i & 0 \\
                           i &  0 & 0 \\
                           0 &  0 & 0
                         \end{array}
                   \Bigg)
\end{equation*}
are the generators of SO(3) rotation about the respective coordinate axes of the LCS and
$\mathbf{w}=\frac{\mathbf k}{k}$
is the unit wavevector.
Substituting Eqs. (\ref{R-LTA1}) and (\ref{QUT1}) into Eq. (\ref{JV'}), one has
\begin{equation}\label{T-JV}
    \alpha' =\exp \left(i \hat{\sigma}_3 \Phi \right) \alpha,
\end{equation}
where the relation
\begin{equation}\label{HO}
    \hat{\sigma}_3
   =\varpi^\dag (\hat{\mathbf \Sigma} \cdot \mathbf{w}) \varpi ,
\end{equation}
which holds irrespective of the SV in $\varpi$, has been used. Equation (\ref{T-JV}) is the transformation of the Jones spinor under the change of the SV.

It is noted that Eq. (\ref{HO}) reflects the correspondence between the SO(3) and SU(2) rotations \cite{Tung}.
In fact, a comparison of Eq. (\ref{T-JV}) with Eq. (\ref{JV'}) reveals
$$ \varpi'^\dag \mathbf{a}= \exp(i \hat{\sigma}_3 \Phi) \varpi^\dag \mathbf{a} $$
when Eq. (\ref{QUT2}) is taken into account. Considering the arbitrariness of the polarization vector $\mathbf a$, one must have
$\varpi'^\dag =\exp(i \hat{\sigma}_3 \Phi) \varpi^\dag$
or, equivalently,
\begin{equation}\label{R-LTA2}
    \varpi' =\varpi \exp(-i \hat{\sigma}_3 \Phi).
\end{equation}
A comparison of it with Eq. (\ref{R-LTA1}) shows that the SO(3) rotation of the transverse axes about the momentum corresponding to the change of the SV can also be expressed by a SU(2) rotation.
More interestingly, the generator of the SU(2) rotation about the momentum does not depend on the momentum. Instead, it is the constant Pauli matrix $\hat{\sigma}_3$.
This in turn means that transformation (\ref{T-JV}) is in fact a SU(2) rotation of the Jones spinor about the momentum.
According to Eqs. (\ref{TA}), the rotation angle $\Phi$ determined by Eq. (\ref{R-LTA1}) or (\ref{R-LTA2}) depends not only on the SV's $ \mathbf{I}' $ and $ \mathbf{I} $ but also on the unit wavevector $\mathbf{w}$ unless
$ \mathbf{I}' =-\mathbf{I} $. In that case, one has $ \Phi=\pi $.

As emphasized before, the polarization vector always depends on the momentum due to the constraint of transversality condition (\ref{TC}). But since the polarization bases in matrix (\ref{varpi}) have already satisfied Eq. (\ref{TC}), the Jones spinor (\ref{QUT2}) is no longer subject to such conditions. That is to say, the Jones spinor can in principle be a constant function. This is one property that distinguishes the Jones spinor from the polarization vector.
However, it can be seen from Eq. (\ref{T-JV}) that even if the unprimed Jones spinor of a polarization state is constant, its primed one is not. Let us analyze what this SU(2) rotation means to the Jones spinor.

\subsection{Jones spinor is defined over NCS}

To this end, we resort to the Stokes parameters in association with the primed SV, which are given by
\begin{equation}\label{SP'}
    s'_i =\alpha'^\dag \hat{\sigma}_i \alpha',
\end{equation}
in accordance with Eq. (\ref{SP}). Upon substituting Eq. (\ref{T-JV}) and using Eq. (\ref{SP}), one gets
\begin{subequations}\label{T-SP}
\begin{align}
  s'_1 & =  s_1 \cos 2\Phi +s_2 \sin 2\Phi,  \label{SP1} \\
  s'_2 & = -s_1 \sin 2\Phi +s_2 \cos 2\Phi,  \label{SP2} \\
  s'_3 & =  s_3.                             \label{SP3}
\end{align}
\end{subequations}
These are the transformations of the Stokes parameters under the change of the SV. Due to correspondence (\ref{HO}), only the first two Stokes parameters change with the SV. The third one does not.
It is noted that when $ \mathbf{I}' =-\mathbf{I} $, that is to say, when $ \Phi=\pi $, one has $ s'_i =s_i $.

The Stokes parameters in the case of a plane wave form the Stokes vector \cite{Coll, Gold, Born-W, Jauc-R, Fano}, which is usually depicted on the surface of the Poincar\'{e} sphere.
But because transformations (\ref{T-SP}) result from a rotation of the NCS about the momentum, the third Stokes parameter that is invariant under such a rotation should be the component of the Stokes vector along the momentum. Accordingly, the first two Stokes parameters in association with one SV should be the components of the Stokes vector along the transverse axes of the corresponding NCS.
Specifically, the unprimed Stokes parameters (\ref{SP}) form the following Stokes vector,
\begin{equation}\label{SV}
    \mathbf{s}=s_1 \mathbf{u} +s_2 \mathbf{v} +s_3 \mathbf{w},
\end{equation}
which as a whole is associated with the unprimed SV.
Letting
\begin{equation}\label{PMV}
\hat{\boldsymbol \sigma}
=\hat{\sigma}_1 \mathbf{u} +\hat{\sigma}_2 \mathbf{v}
+\hat{\sigma}_3 \mathbf{w},
\end{equation}
one can rewrite Eq. (\ref{SV}) as
\begin{equation}\label{SV-PMV}
\mathbf{s}=\alpha^\dag \hat{\boldsymbol \sigma} \alpha.
\end{equation}
It states that the Jones spinor (\ref{QUT2}) in association with one particular SV via Eqs. (\ref{varpi}) and (\ref{TA}) is defined over the NCS that is determined by that same SV.
This is another property that distinguishes the Jones spinor from the polarization vector, which is defined over the LCS.

To demonstrate this in detail, we make use of the primed Stokes vector given by
\begin{equation}\label{SV'}
    \mathbf{s}'=\alpha'^\dag \hat{\boldsymbol \sigma}' \alpha'
               =s'_1 \mathbf{u}' +s'_2 \mathbf{v}' +s'_3 \mathbf{w},
\end{equation}
where
$\hat{\boldsymbol \sigma}' =\hat{\sigma}_1 \mathbf{u}' +\hat{\sigma}_2 \mathbf{v}' +\hat{\sigma}_3 \mathbf{w}$.
With the help of Eqs. (\ref{T-SP}) and (\ref{R-TA}), one readily finds that it is related to the unprimed Stokes vector (\ref{SV}) via
\begin{equation}\label{T-SV}
    \mathbf{s}'
   =\exp [i(\hat{\mathbf \Sigma} \cdot \mathbf{w}) \Phi] \mathbf{s}.
\end{equation}
This is explained in terms of their transverse components
$\mathbf{s}'_\perp =s'_1 \mathbf{u}' +s'_2 \mathbf{v}'$
and
$\mathbf{s}_\perp =s_1 \mathbf{u} +s_2 \mathbf{v}$
as follows.
As is well known, the Pauli matrices (\ref{PM}) satisfy the SU(2) algebra,
\begin{equation}\label{CCR}
    \Big[ \frac{\hat{\sigma}_i}{2}, \frac{\hat{\sigma}_j}{2} \Big]
   =i \sum_k \varepsilon_{ijk} \frac{\hat{\sigma}_k}{2},
\end{equation}
where $\varepsilon_{ijk}$ is the Levi-Civit\'{a} pseudotensor. In view of this, it follows from Eq. (\ref{PMV}) that each of them, except for a factor $\frac{1}{2}$, is the generator of a SU(2) rotation about the local Cartesian axes of the NCS.
The primed Jones spinor is therefore the result of the rotation of the unprimed Jones spinor about the momentum by an angle $-2 \Phi$ as is shown by Eq. (\ref{T-JV}).
If the transverse axes of the NCS were not rotated, the first two primed Stokes parameters $s'_1$ and $s'_2$ would form a transverse component,
$\mathbf{s}''_\perp =s'_1 \mathbf{u} +s'_2 \mathbf{v}$,
which is equal to the result of the rotation of $\mathbf{s}_\perp$ about the momentum by the same angle. This is what Eqs. (\ref{SP1}) and (\ref{SP2}) show and is graphically displayed in Fig. 1(a).
\begin{figure}[tb]
	\centerline{\includegraphics[width=11cm]{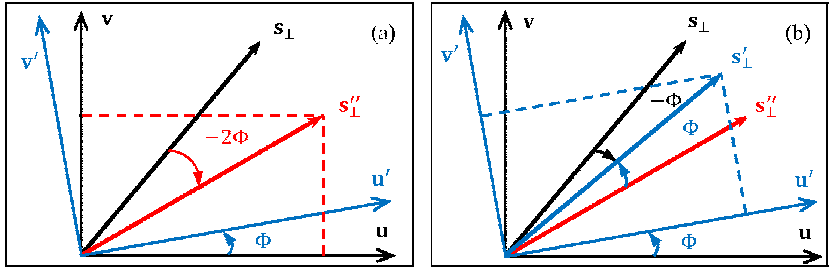}}
	\caption{Relation between $\mathbf{s}'_\perp$ and $\mathbf{s}_\perp$. (a) $\mathbf{s}''_\perp$ is the result of the rotation of $\mathbf{s}_\perp$ by an angle $-2 \Phi$ within the unprimed NCS. (b) $\mathbf{s}''_\perp$ is rotated to $\mathbf{s}'_\perp$ along with the unprimed NCS being rotated to the primed NCS by an angle $\Phi$.}
\end{figure}
However, as mentioned above, the transverse axes $\mathbf{u}'$ and $\mathbf{v}'$ of the primed NCS result from the rotation of the transverse axes $\mathbf{u}$ and $\mathbf{v}$ of the unprimed NCS about the momentum by the angle $\Phi$.
Along with the unprimed NCS being rotated to the primed one, the transverse component $\mathbf{s}''_\perp$ is rotated to $\mathbf{s}'_\perp$ as is displayed in Fig. 1(b). As a consequence, $\mathbf{s}'_\perp$ is equal to the result of the rotation of $\mathbf{s}_\perp$ about the momentum by an angle $-\Phi$. This is just what Eq. (\ref{T-SV}) means.
We are thus convinced that the Jones spinor (\ref{QUT2}) in association with one particular SV is defined over the NCS that is determined exactly by that SV.
The Stokes parameters it gives are quantities with respect to the same NCS, forming the Stokes vector (\ref{SV}).
By the way, it is pointed out that the primed Stokes vector (\ref{SV'}) is in general different from the unprimed one (\ref{SV}) even when
$ \mathbf{I}' =-\mathbf{I} $.

\subsection{SV fixes a natural representation for polarization}

We are now in a position to discuss how the SV-associated Stokes parameters characterize the polarization state of light beams.
It has been shown that matrix (\ref{varpi}) obeys Eq. (\ref{QU1}). With the help of this property, one readily gets 
$
\mathbf{a}^\dag \mathbf{a}=\alpha^\dag \alpha
$
from Eq. (\ref{QUT1}).
Upon substituting Eq. (\ref{QUT2}) and considering the arbitrariness of the polarization vector $\mathbf a$ into account, one immediately arrives at
\begin{equation}\label{QU2}
    \varpi \varpi^\dag=I_3 ,
\end{equation}
where $I_3$ is the 3-by-3 unit matrix.
Equations (\ref{QU1}) and (\ref{QU2}) indicate that matrix (\ref{varpi}) in association with the SV through Eqs. (\ref{TA}) is a quasi-unitary matrix \cite{Golu-L}. $\varpi^\dag$ is the Moore-Penrose pseudo inverse of $\varpi$, and vice versa.
This means that the SV fixes a quasi-unitary transformation between the polarization vector over the LCS and the Jones spinor over the associated NCS.
Specifically, to each polarization vector over the LCS there corresponds one unique Jones spinor over the NCS that is determined by a SV, and vice versa.
In the language of quantum mechanics, the SV fixes, via Eqs. (\ref{varpi}) and (\ref{QUT2}), a representation for the polarization.
Remarkably, this is a representation that is distinguished from the laboratory representation in which the state of polarization is represented by the polarization vector over the LCS.
It is a natural representation.
The polarization wavefunction in such a representation is the Jones spinor over the associated NCS. It describes the polarization state of light beams in such a way that the Stokes parameters it gives through the Pauli matrices are quantities with respect to that same NCS. In a word, only in a particular natural representation can the Stokes parameters characterize the polarization state of light beams.

More importantly, because matrix (\ref{varpi}) is quasi unitary no matter what the SV is, there are an infinite number of natural representations for the polarization. Each SV will fix a different natural representation.
On one hand, all the natural representations are equivalent in the sense that a polarization state can be described in any natural representation by a corresponding wavefunction. The wavefunctions of a polarization state in different natural representations are not the same as is shown by Eq. (\ref{T-JV}). As a consequence, according to Eqs. (\ref{T-SP}), the Stokes parameters of a polarization state with respect to different NCS's are in general not the same, either.
The situation encountered here is similar to what is encountered in Einstein's theory of special relativity in which the momentum and energy of a particle depend on the choice of inertial reference frame \cite{Gold1}.
On the other hand, it is seen from Eq. (\ref{QUT1}) that a Jones spinor in different natural representations describes different states of polarization.
For instance, the polarization state described by the same Jones spinor $\alpha$ in the primed natural representation will be given by
$ \mathbf{a}' =\varpi' \alpha $.
Upon substituting Eq. (\ref{R-LTA1}) and making use of Eq. (\ref{QUT1}), one has
\begin{equation}\label{RoPV}
    \mathbf{a}' =\exp [-i(\hat{\mathbf \Sigma} \cdot \mathbf{w}) \Phi] \mathbf{a}.
\end{equation}
In this sense, all the natural representations are not equivalent. Let us explain the underlying reason below.

\section{\label{SV-perp} Two independent degrees of freedom of polarization}

As emphasized before, the Jones spinor is not constrained by such conditions as Eq. (\ref{TC}). It can in principle be a constant function.
We know \cite{Mess, Merz} that a constant spinor in quantum mechanics describes the intrinsic degree of freedom of the electron, the spin. The operator for the spin in such a representation is the Pauli matrices.
With that in mind, we have reason to believe that a constant Jones spinor in the natural representation describes such a degree of freedom of a light beam that is represented by the Pauli matrices (\ref{PM}).
But what is unexpected is that it is not a degree of freedom with respect to the LCS. It is a degree of freedom with respect to the NCS as can be seen from Eq. (\ref{SV-PMV}). The term \emph{intrinsic} is deemed more appropriate for it.
By this it is meant that the polarization of a light beam is not the same as its intrinsic degree of freedom. To characterize the polarization of a light beam with its intrinsic degree of freedom, one has to know how the NCS is related to the LCS. This is what Eq. (\ref{QUT1}) implies.
In other words, the SV that determines the transverse axes of the NCS through Eqs. (\ref{TA}) shows up as another degree of freedom to characterize the polarization of a light beam.
In a word, an exact characterization of the polarization of a light beam needs these two kinds of degrees of freedom.
It should be noted, however, that different from the intrinsic degree of freedom, the SV is a new kind of degree of freedom. It is not an hermitian operator per se.

From this result it is inferred that a definite value of the intrinsic degree of freedom in combination with different SV's will mean different states of polarization.
To illustrate this with examples, we examine in this section the meaning of the eigenvalue of the Pauli matrices in combination with different SV's that are perpendicular to the $z$ axis.
As a first example, we consider the Pauli matrix $\hat{\sigma}_1$, which has eigenfunctions
$
\alpha_{1+}=\Big(\begin{array}{c}
                   1 \\
                   0
                 \end{array}
            \Big)
$
and
$
\alpha_{1-}=\Big(\begin{array}{c}
                   0 \\
                   1
                 \end{array}
            \Big)
$
with eigenvalues $\sigma_{1+} = +1$ and $\sigma_{1-} = -1$, respectively.
Let us start with the perpendicular SV along the $x$ axis, denoted by
$\mathbf{I}^{\perp} =\bar{x}$.
The transverse axes of the NCS it determines take the form
\begin{subequations}\label{TA-perp}
\begin{align}
  \mathbf{u}^{\perp} =& \frac{1}{C}
                   \Big(\bar{x}-\bar{\rho} \frac{k_\rho^2}{k^2} \cos \varphi
                        -\bar{z} \frac{k_z k_\rho}{k^2} \cos \varphi \Big), \\
  \mathbf{v}^{\perp} =& \frac{1}{C} \Big(\bar{y} \frac{k_z}{k}
                                   -\bar{z} \frac{k_\rho}{k} \sin \varphi \Big),
\end{align}
\end{subequations}
where $\bar y$ and $\bar z$ are unit vectors along the $y$ and $z$ axes, respectively,
$\bar \rho$ stands for the radial unit vector in cylindrical coordinates,
$k_z =(k^2-k^2_\rho)^{1/2}$ is the axial component of $\mathbf k$, and
$
C=\big(1-\frac{k_\rho^2}{k^2} \cos^2 \varphi \big)^{1/2}.
$
In this case, the polarization states described by $\alpha_{1+}$ and $\alpha_{1-}$ are given by
\begin{equation}\label{PV-UP1}
    \mathbf{a}_{1+}^{\perp} = \varpi^{\perp} \alpha_{1+} =\mathbf{u}^{\perp}, \quad
    \mathbf{a}_{1-}^{\perp} = \varpi^{\perp} \alpha_{1-} =\mathbf{v}^{\perp},
\end{equation}
respectively, where
$\varpi^{\perp} =(\begin{array}{cc}
                    \mathbf{u}^{\perp} & \mathbf{v}^{\perp}
                  \end{array}
                 )$.
They are mutually orthogonal.
When the perpendicular SV $\mathbf{I}^\perp$ is rotated to $\mathbf{I}'^{\perp}= \bar{y}$ about the $z$ axis, the transverse axes of the NCS become
\begin{eqnarray*}
  \mathbf{u}'^{\perp} &=& \frac{1}{C'}
                   \Big(\bar{y}\frac{k^2_z}{k^2}+\bar{\varphi} \frac{k_\rho^2}{k^2} \cos \varphi
                        -\bar{z} \frac{k_z k_\rho}{k^2} \sin \varphi \Big), \\
  \mathbf{v}'^{\perp} &=& \frac{1}{C'} \Big(-\bar{x} \frac{k_z}{k}
                                    +\bar{z} \frac{k_\rho}{k} \cos \varphi \Big),
\end{eqnarray*}
where $\bar{\varphi}$ stands for the azimuthal unit vector in cylindrical coordinates and
$
C'=(1-\frac{k_\rho^2}{k^2} \sin^2 \varphi )^{1/2}.
$
Now the polarization states described by $\alpha_{1+}$ and $\alpha_{1-}$ are given by
\begin{equation}\label{PV-UP2}
    \mathbf{a}'^{\perp}_{1+} = \varpi'^{\perp} \alpha_{1+} =\mathbf{u}'^{\perp}, \quad
    \mathbf{a}'^{\perp}_{1-} = \varpi'^{\perp} \alpha_{1-} =\mathbf{v}'^{\perp},
\end{equation}
respectively, where
$\varpi'^{\perp} =(\begin{array}{cc}
                    \mathbf{u}'^{\perp} & \mathbf{v}'^{\perp}
                  \end{array}
                 )$.
Clearly, the polarization state $\mathbf{a}'^{\perp}_{1+}$ ($\mathbf{a}'^{\perp}_{1-}$) is different from $\mathbf{a}^{\perp}_{1+}$ ($\mathbf{a}^{\perp}_{1-}$).
In addition, even though $\mathbf{I}'^\perp$ results from the rotation of $\mathbf{I}^{\perp}$ about the $z$ axis, the polarization vector $\mathbf{a}'^{\perp}_{1+}$ ($\mathbf{a}'^{\perp}_{1-}$) cannot be obtained from $\mathbf{a}^{\perp}_{1+}$ ($\mathbf{a}^{\perp}_{1-}$) by the same rotation.
As a matter of fact, from Eq. (\ref{RoPV}) it follows that
\begin{equation*}
    \mathbf{a}'^{\perp}_{1+} = \exp [-i(\hat{\mathbf \Sigma} \cdot \mathbf{w}) \Phi^\perp ]
                               \mathbf{a}_{1+}^{\perp},                                    \quad
    \mathbf{a}'^{\perp}_{1-} = \exp [-i(\hat{\mathbf \Sigma} \cdot \mathbf{w}) \Phi^\perp ]
                               \mathbf{a}_{1-}^{\perp},
\end{equation*}
where $\Phi^\perp$ is the local rotation angle satisfying
\begin{equation}\label{RoP-P}
    \varpi'^\perp =\varpi^\perp \exp(-i \hat{\sigma}_3 \Phi^\perp),
\end{equation}
in accordance with Eq. (\ref{R-LTA2}).
Generally speaking, any rotation of $\mathbf{I}^{\perp}$ about the $z$ axis will change the polarization states (\ref{PV-UP1}).

Clearly, whether polarization vectors (\ref{PV-UP1}) or polarization vectors (\ref{PV-UP2}) have first-order axial components with respect to $k_{\rho}/k$.
If the strength factor $e(\mathbf{k})$ is sharply peaked at $k_{\rho}=0$, a paraxial approximation can be made to them when Eqs. (\ref{FT}) and (\ref{vec-e}) are taken into consideration. In the zeroth-order approximation in which $\frac{k_\rho}{k} \approx 0$ and $\frac{k_z}{k} \approx 1$,
polarization vectors (\ref{PV-UP1}) reduce to
\begin{equation*}
    \mathbf{a}_{1+}^{\perp} \approx \bar x, \quad
    \mathbf{a}_{1-}^{\perp} \approx \bar y.
\end{equation*}
They will give rise to light beams of homogeneous $x$ and $y$ ``polarizations'', respectively. In the same approximation, polarization vectors (\ref{PV-UP2}) reduce to
\begin{equation*}
    \mathbf{a}'^{\perp}_{1+} \approx  \bar{y}, \quad
    \mathbf{a}'^{\perp}_{1-} \approx -\bar{x}.
\end{equation*}
They will also give rise to homogeneous ``polarizations'', but in $y$ and minus $x$ directions, respectively.
It should be pointed out that the polarization vector $\mathbf{a}'^{\perp}_{1+}$ in (\ref{PV-UP2}) and the polarization vector $\mathbf{a}^{\perp}_{1-}$ in (\ref{PV-UP1}) are different though they tend to the same form $\bar y$ in the paraxial approximation. The same is true of the relationship between the polarization vector $\mathbf{a}'^{\perp}_{1-}$ in (\ref{PV-UP2}) and the polarization vector $\mathbf{a}^{\perp}_{1+}$ in (\ref{PV-UP1}).
From these discussions we are led to a surprising conclusion that the polarization state of a paraxial beam that is expressed only by the polarization vector $\bar x$ or $\bar y$ is not yet completely specified.
By this it is meant that the exact meaning of the so-called horizontal or vertical ``polarization'' in the literature \cite{Aiel-TMGL, Fick-LRZ, Kwia-MWZ, Mich-WZ, Kies-SWUW, Simo-SG, Pan-CL, Bayr-SCB, Ecker} is not clear.

The meaning of the eigenvalue of the Pauli matrix $\hat{\sigma}_2$ can be discussed similarly. More interesting is the meaning of the eigenvalue of the Pauli matrix $\hat{\sigma}_3$ because of correspondence (\ref{HO}).
The eigenfunction of $\hat{\sigma}_3$ with eigenvalue $\sigma_{3}= \pm 1$ assumes
$
\alpha_{\sigma_3}= \frac{1}{\sqrt 2} \Big(\begin{array}{c}
                                                    1 \\
                                                    i \sigma_3
                                                  \end{array}
                                             \Big)
$,
satisfying
\begin{equation}\label{EVE}
    \hat{\sigma}_3 \alpha_{\sigma_3} =\sigma_3 \alpha_{\sigma_3}.
\end{equation}
If the perpendicular SV is along the $x$ axis, $\mathbf{I}^{\perp} =\bar{x}$, the polarization state it describes is given by
\begin{equation}\label{ESoH}
    \mathbf{a}^\perp_{\sigma_3}
   =\varpi^\perp \alpha_{\sigma_3}
   =\frac{1}{\sqrt 2}(\mathbf{u}^\perp +i \sigma_3 \mathbf{v}^\perp ),
\end{equation}
where $\mathbf{u}^\perp$ and $\mathbf{v}^\perp$ are the polarization bases (\ref{TA-perp}).
When the perpendicular SV $\mathbf{I}^\perp$ is rotated to $\mathbf{I}'^{\perp} =\bar{y}$ about the $z$ axis, the polarization state it describes becomes
\begin{equation}\label{ESoH'}
    \mathbf{a}'^\perp_{\sigma_3}
   =\varpi'^\perp \alpha_{\sigma_3}
   =\frac{1}{\sqrt 2}( \mathbf{u}'^\perp +i \sigma_3 \mathbf{v}'^\perp ).
\end{equation}
The same as before, polarization state (\ref{ESoH'}) cannot be the same as polarization state (\ref{ESoH}). Moreover, the former cannot be obtained through a rotation of the latter about the $z$ axis.
But on the other hand, it is easy to show with the help of Eqs. (\ref{RoP-P}) and (\ref{EVE}) that they are related to each other by a $\sigma_3$-dependent phase factor,
\begin{equation}\label{BP}
    \mathbf{a}'^\perp_{\sigma_3}=\exp(-i \sigma_3 \Phi^\perp ) \mathbf{a}^\perp_{\sigma_3}.
\end{equation}
This result can be understood as follows.
Considering Eqs. (\ref{QU1}) and (\ref{QU2}) into account, one readily gets from Eq. (\ref{HO})
\begin{equation*}
    \hat{\mathbf \Sigma} \cdot \mathbf{w}
   =\varpi \hat{\sigma}_3 \varpi^\dag .
\end{equation*}
Because this correspondence holds irrespective of the SV in the transformation matrix $\varpi$, polarization states (\ref{ESoH}) and (\ref{ESoH'}) are both the eigenstates of $\hat{\mathbf \Sigma} \cdot \mathbf{w}$ with the same eigenvalue,
\begin{equation*}
    (\hat{\mathbf \Sigma} \cdot \mathbf{w}) \mathbf{a}^\perp_{\sigma_3}
   =\sigma_3 \mathbf{a}^\perp_{\sigma_3},                              \quad
    (\hat{\mathbf \Sigma} \cdot \mathbf{w}) \mathbf{a}'^\perp_{\sigma_3}
   =\sigma_3 \mathbf{a}'^\perp_{\sigma_3}.
\end{equation*}

In the zeroth-order paraxial approximation, polarization vector (\ref{ESoH}) reduces to
\begin{equation}\label{ESoH-A}
    \mathbf{a}^\perp_{\sigma_3}
    \approx \frac{1}{\sqrt 2} (\bar{x} +i \sigma_3 \bar{y}).
\end{equation}
It will give rise to light beams of homogeneous ``circular polarization'' as can be seen from Eqs. (\ref{FT}) and (\ref{vec-e}).
In the meanwhile, polarization vector (\ref{ESoH'}) reduces to
\begin{equation}\label{ESoH'-A}
    \mathbf{a}'^\perp_{\sigma_3}
    \approx \frac{\exp ( -i \sigma_3 \frac{\pi}{2} )}{\sqrt 2}
             (\bar{x} +i \sigma_3 \bar{y}),
\end{equation}
which will also give rise to light beams of homogeneous ``circular polarization''. The paraxially approximated polarization vectors (\ref{ESoH-A}) and (\ref{ESoH'-A}) are both the eigenfunctions of $\hat{\mathbf \Sigma} \cdot \bar{z}$ with the same eigenvalue $\sigma_3$. But as has been shown above, they do not represent the same polarization state.
The $\frac{\pi}{2}$ in the phase factor of expression (\ref{ESoH'-A}) now conveys the rotation angle of the perpendicular SV from $\mathbf{I}^\perp $ to $\mathbf{I}'^\perp $ about the $z$ axis.
From these discussions we are led to a conclusion that the polarization state of a paraxial beam that is expressed only by the polarization vector
$\frac{1}{\sqrt 2} (\bar{x} +i \sigma_3 \bar{y})$
is not yet completely specified.

\section{\label{no-entangle} Vector vortex beams do not have any entanglement}

Now we are ready to demonstrate that vector vortex beams are not endowed with any entanglement. For this purpose, we consider a SV along the minus $z$ direction, denoted by
$\mathbf{I}^\parallel = -\bar z$.
The transverse axes of the NCS it determines are given by
\begin{eqnarray*}
  \mathbf{u}^\parallel &=& \frac{k_z}{k} \bar{\rho}-\frac{k_\rho}{k} \bar{z}, \\
  \mathbf{v}^\parallel &=& \bar{\varphi}. 
\end{eqnarray*}
In this case, the polarization states that are described by $\alpha_{1+}$ and $\alpha_{1-}$, the eigenfunctions of the Pauli matrix $\hat{\sigma}_1$, are given by
\begin{subequations}\label{PV-VVB}
\begin{align}
  \mathbf{a}_{1+}^\parallel =\varpi^\parallel \alpha_{1+} &
 =\frac{k_z}{k} \bar{\rho}-\frac{k_\rho}{k} \bar{z},               \\
  \mathbf{a}_{1-}^\parallel =\varpi^\parallel \alpha_{1-} &
 =\bar{\varphi},
\end{align}
\end{subequations}
respectively, where
$\varpi^\parallel =(\begin{array}{cc}
              \mathbf{u}^\parallel & \mathbf{v}^\parallel
            \end{array}
           )$.
It is noted that the polarization vector $\mathbf{a}_{1-}^\parallel$ has no axial component whether the paraxial approximation is made or not.
Note also that different from polarization vectors (\ref{PV-UP1}) or (\ref{PV-UP2}) that are regular on the propagation axis, $k_\rho =0$, polarization vectors (\ref{PV-VVB}) are indeterminate on the propagation axis.

To obtain an explicit expression for vector vortex beams, we consider the following strength factor,
\begin{equation}\label{OAMB}
    e(\mathbf{k})= i^{-1} f(k_\rho) \exp(im \varphi),
\end{equation}
where $f(k_\rho)$ is any physically allowed function, $m$ is an integer, and the additional factor $i^{-1}$ is introduced for later convenience.
Substituting Eq. (\ref{vec-e}) into Eq. (\ref{FT}) and making use of Eqs. (\ref{PV-VVB}) and (\ref{OAMB}), we find that the electric fields produced by polarization vectors $\mathbf{a}_{1+}^\parallel$ and $\mathbf{a}_{1-}^\parallel$ are given by
\begin{subequations}\label{VVB}
\begin{align}
  \mathbf{E}^\parallel_{1+} &
 =\frac{1}{2} i^m
   [(A_{m+1}-A_{m-1}) \bar{r} -i (A_{m+1}+A_{m-1}) \bar{\phi} +2i B_m \bar{z}]
   e^{i m \phi}, \label{VVB1}                                                    \\
  \mathbf{E}^\parallel_{1-} &
 =\frac{1}{2} i^m [(A'_{m+1}-A'_{m-1}) \bar{\phi} +i (A'_{m+1}+A'_{m-1}) \bar{r}]
   e^{i m \phi}, \label{VVB2}
\end{align}
\end{subequations}
respectively, where
\begin{eqnarray*}
  A_m (r,z) &=& \int_0^k \frac{k_z}{k} f(k_\rho) J_m (k_\rho r) e^{i k_z z} k_\rho d k_\rho, \\
  B_m (r,z) &=& \int_0^k \frac{k_\rho}{k} f(k_\rho) J_m (k_\rho r) e^{i k_z z} k_\rho d k_\rho,\\
  A'_m (r,z) &=& \int_0^k f(k_\rho) J_m (k_\rho r) e^{i k_z z} k_\rho d k_\rho,
\end{eqnarray*}
the identity
\begin{equation*}
    \exp(i \xi \cos \psi)= \sum_{m=- \infty}^{\infty} i^m J_m (\xi) \exp (i m \psi)
\end{equation*}
has been used,
$r$, $\phi$, and $z$ are cylindrical coordinates in position space, $\bar r$ and $\bar \phi$ are radial and azimuthal unit vectors, respectively,
and $J_m $ is the Bessel function of the first kind. They are mutually orthogonal vector vortex beams. This shows that the parallel SV $\mathbf{I}^\parallel$ is well suited to characterize the polarization state of vector vortex beams.
It is worth reminding that, in free space, the principle of duality applies \cite{Lawr-P}. So vector vortex beam (\ref{VVB1}) is a TM mode and vector vortex beam (\ref{VVB2}) a TE mode.
When $m=0$, Eqs. (\ref{VVB}) reduce to the following ``radially and azimuthally polarized'' cylindrical-vector beams,
\begin{equation*}
    \mathbf{E}^\parallel_{1+} =[A_1 \bar{r} +i B_0 \bar{z}], \quad
    \mathbf{E}^\parallel_{1-} =A'_1 \bar{\phi},
\end{equation*}
respectively.
Corresponding to the indeterminacy of polarization vectors (\ref{PV-VVB}) on the propagation axis, the unit vectors $\bar r$ and $\bar \phi$ in the above expressions are also indeterminate on the propagation axis, $r=0$. But meanwhile, the scalar functions before them, $A_1$ and $A'_1$, both vanish on the propagation axis no matter what the function $f(k_\rho)$ in the strength factor (\ref{OAMB}) is. This indicates that there is no problem with the SV-rooted indeterminacy of the polarization bases (\ref{TA}).

With definite values of the SV and intrinsic degree of freedom, the polarizations of vector vortex beams (\ref{VVB}) do not have any entanglement with other degrees of freedom.
Interestingly, Eq. (\ref{VVB2}) can be rewritten as
\begin{equation}\label{TE}
    \mathbf{E}^\parallel_{1-}
   =\frac{1}{2} i^{m+1}
    [A'_{m+1} e^{i \phi}(\bar{x}-i \bar{y}) +A'_{m-1} e^{-i \phi}(\bar{x}+i \bar{y})]
     e^{i m \phi}.
\end{equation}
It is on the basis of such an expression that the TE vortex beam was considered \cite{McLa-KF, D'Am-CG} as entangled between polarization and orbital angular momentum.
We are now clear that it is incorrect to consider the mutually orthogonal unit vectors
$\frac{1}{\sqrt 2}(\bar{x}+i \bar{y})$ and $\frac{1}{\sqrt 2}(\bar{x}-i \bar{y})$
as the polarization bases of the beam (\ref{TE}).
After all, the SV of its polarization is parallel, rather than perpendicular, to the $z$ axis.
Moreover, it is not a paraxially approximated beam.
The same discussions apply also to TM vortex beam (\ref{VVB1}).

As a matter of fact, apart from having definite states of polarization, vector vortex beams (\ref{VVB}) also both have definite orbital angular momenta in the $z$ direction.
To understand this, it is just mentioned that in any natural representation, the orbital-angular-momentum operator $\hat{l}_z =-i \frac{\partial}{\partial \varphi}$ is independent of the Pauli matrices, the intrinsic degree of freedom of the polarization.
In view of this, it follows from Eq. (\ref{OAMB}) that vector vortex beams (\ref{VVB}) are both the eigenstates of $\hat{l}_z$ with the same eigenvalue $m$. Further discussions \cite{Li15} are beyond the scope of present paper.

\section{\label{CandR} Conclusions and remarks}

In summary, we found that an exact characterization of the polarization state of light beams needs two different kinds of degrees of freedom.
One is the SV that determines through Eqs. (\ref{TA}) how the transverse axes of the NCS are related to the LCS. From the point of view of quantum mechanics, it fixes a natural representation for the polarization in which the wavefunction of the polarization, the Jones spinor (\ref{QUT2}), is defined over the associated NCS. 
The other is their intrinsic degree of freedom with respect to the NCS, which is represented by the Pauli matrices (\ref{PM}).
In other word, the polarization of light beams conveys how their intrinsic degree of freedom with respect to the NCS behaves in the LCS.
On one hand, all the natural representations are equivalent in the sense that a polarization state can be described in any natural representation by a corresponding Jones spinor. The Stokes parameters of a polarization state given by its Jones spinor in a particular natural representation are quantities with respect to the associated NCS.
So the Stokes parameters alone are not enough to completely characterize the polarization state of light beams.
On the other hand, all the natural representations are not equivalent in the sense that a Jones spinor in different natural representations does not describe the same polarization state. As a result, a definite value of the intrinsic degree of freedom with respect to different NCS's manifests as different states of polarization in the LCS.
From this result it follows that the exact meaning of the so-called horizontal or vertical ``polarization'' in the paraxial approximation is unclear. The same is true of the so-called circular ``polarization'' in the paraxial approximation.
On the basis of these discussions we showed that the representative vector vortex beams expressed by Eqs. (\ref{VVB}) are not entangled in polarization and orbital angular momentum.

What underlies the presented properties of the polarization of light beams is the constraint of transversality condition (\ref{TC}).
Firstly, it implies that represented by the polarization vector, the notion of polarization cannot be a single degree of freedom.
Secondly, the fact that the Stokes parameters at each momentum are physical quantities with respect to the NCS at that momentum is in consistency with the constraint that the polarization vector is orthogonal to the associated momentum.
However, it should be remarked that these properties could not be formulated within the framework of Maxwell's classical theory although the transverse nature of the electric field is correctly expressed by the Maxwell equation (\ref{ME}).
This is because the polarization is ultimately quantized as is explicitly indicated by the SU(2) commutation relation (\ref{CCR}).
It is also remarked that the identification of the natural representation may indicate that the quantization of the polarization is a local phenomenon in momentum space.
The findings pave the way towards genuine quantization of the polarization of light beams.

\begin{acknowledgments}

This work was supported in part by the program of Shanghai Municipal Science and Technology Commission under Grant 18ZR1415500.

\end{acknowledgments}



\begin{thebibliography}{99}

\bibitem{Youn-B}
K. S. Youngworth and T. G. Brown, Focusing of high numerical aperture cylindrical-vector beams, Opt. Express {\bf 7}, 77-87 (2000).

\bibitem{Zhan}
Q. Zhan, Cylindrical Vector Beams: from Mathematical Concepts to Applications, Adv. Opt. Photon. {\bf 1}, 1-57 (2009).

\bibitem{Topp-AMGL}
F. T\"{o}ppel, A. Aiello, C. Marquardt, E. Giacobino, and G. Leuchs, Classical entanglement in polarization metrology, New J. Phys. {\bf 16}, 073019 (2014).

\bibitem{Aiel-TMGL}
A. Aiello, F. T\"{o}ppel, C. Marquardt, E. Giacobino, and G. Leuchs, Quantum-like nonseparable structures in optical beams, New J. Phys. {\bf 17}, 043024 (2015).

\bibitem{Mili-NLNA}
G. Milione, T. An Nguyen, J. Leach, D. A. Nolan, and R. R. Alfano, Using the nonseparability of vector beams to encode information for optical communication, Opt. Lett. {\bf 40}, 4887-4890 (2015).

\bibitem{McLa-KF}
M. McLaren, T. Konrad, and A. Forbes, Measuring the nonseparability of vector vortex beams, Phys. Rev. A {\bf 92}, 023833 (2015).

\bibitem{D'Am-CG}
V. D'Ambrosio, G. Carvacho, F. Graffitti, C. Vitelli, B. Piccirillo, L. Marrucci, and F. Sciarrino, Entangled vector vortex beams, Phys. Rev. A {\bf 94}, 030304(R) (2016).

\bibitem{Stal-S}
M. Stalder and M. Schadt, Linearly polarized light with axial symmetry generated by liquid-crystal polarization converters, Opt. Lett. {\bf 21}, 1948-1950 (1996).

\bibitem{Fick-LRZ}
R. Fickler, R. Lapkiewicz, S. Ramelow, and A. Zeilinger, Quantum Entanglement of Complex Photon Polarization Patterns in Vector Beams, Phys. Rev. A {\bf 89}, 060301(R) (2014).

\bibitem{Kwia-MWZ}
P. G. Kwiat, K. Mattle, H. Weinfurter, A. Zeilinger, A. V. Sergienko, and Y. Shih, New High-Intensity Source of Polarization-Entangled Photon Pairs, Phys. Rev. Lett. {\bf 75}, 4337-4341 (1995).

\bibitem{Mich-WZ}
M. Michler, H. Weinfurter, and M. Żukowski, Experiments towards Falsification of Noncontextual Hidden Variable Theories, Phys. Rev. Lett. {\bf 84}, 5457-5461 (2000).

\bibitem{Kies-SWUW}
N. Kiesel, C. Schmid, U. Weber, R. Ursin, and H. Weinfurter, Phys. Rev. Lett. {\bf 95}, 210505 (2005).

\bibitem{Simo-SG}
B. N. Simon, S. Simon, F. Gori, M. Santarsiero, R. Borghi, N. Mukunda, and R. Simon, Nonquantum Entanglement Resolves a Basic Issue in Polarization Optics, Phys. Rev. Lett. {\bf 104}, 023901 (2010).

\bibitem{Pan-CL}
J.-W. Pan, Z.-B. Chen, C.-Y. Lu, H. Weinfurter, A. Zeilinger, and M. \.{Z}ukowski, Multiphoton Entanglement and Interferometry, Rev. Mod. Phys. {\bf 84}, 777-838 (2012).

\bibitem{Bayr-SCB}
\"{O}. Bayraktar, M. Swillo, C. Canalias, and G. Bj\"{o}rk, Quantum-polarization State Tomography, Phys. Rev. A {\bf 94}, 020105(R) (2016).

\bibitem{Ecker}
S. Ecker et. al., Overcoming Noise in Entanglement Distribution, Phys. Rev. X {\bf 9}, 041042 (2019).


\bibitem{Meji-MPM}
P.M. Mej\'{\i}as, R. Mart\'{\i}nez-Herrero, G. Piquero, J.M. Movilla, Parametric characterization of the spatial structure of non-uniformly polarized laser beams, Prog. Quant. Electron. {\bf 26}, 65-130 (2002).

\bibitem{Beck-BA}
A. M. Beckley, T. G. Brown, and M. A. Alonso, Full Poincar\'{e} Beams, Opt. Express {\bf 18}, 10777-10785 (2010).

\bibitem{Card}
F. Cardano, E. Karimi, S. Slussarenko, L. Marrucci, C. de Lisio, and E. Santamato, Polarization pattern of vector vortex beams generated by q-plates with different topological charges, Appl. Opt. {\bf 51}, C1-C6 (2012).

\bibitem{Coll}
E. Collett, {\it Field Guide to Polarization} (Bellingham, WA: SPIE, 2005).

\bibitem{Gold}
D. H. Goldstein, {\it Polarized Light}, 3rd ed. (Taylor and Francis, New York, 2011).

\bibitem{Born-W}
M. Born and E. Wolf, {\it Principles of Optics}, 7th ed. (Cambridge University Press, Cambridge, 1999).

\bibitem{Jauc-R}
J. M. Jauch and F. Rohrlich, {\it The Theory of Photons and Electrons}, 2nd ed. (Springer-Verlag, New York, 1976).

\bibitem{Fano}
U. Fano, A Stokes-Parameter Technique for the Treatment of Polarization in Quantum Mechanics, Phys. Rev. {\bf 93}, 121-123 (1954).

\bibitem{Jones}
R. C. Jones, A New Calculus for the Treatment of Optical Systems I: Description and Discussion of the Calculus, J. Opt. Soc. Am. {\bf 31}, 488-493 (1941).

\bibitem{Cald}
C. D. Caldwell, Digital Lock-in Technique for Measurement of Polarization of Radiation, Opt.
Lett. {\bf 1}, 101-103 (1977).

\bibitem{Berr-GL}
H. G. Berry, G. Gabrielse, and A. E. Livingston, Measurement of the Stokes Parameters of
Light, Appl. Opt. {\bf 16}, 3200-3205 (1977).

\bibitem{Gold1}
H. Goldstein, {\it Classical Mechanics}, 2nd ed. (Addison-Wesley, Massachusetts, 1980).

\bibitem{Stra}
J. A. Stratton, {\it Electromagnetic Theory} (McGraw-Hill, New York, 1941).

\bibitem{Gree-W}
H. S. Green and E. Wolf, A Scalar Representation of Electromagnetic Fields, Proc. Phys. Soc. A {\bf 66}, 1129-1137 (1953).

\bibitem{Patt-A}
D. N. Pattanayak and G. P. Agrawal, Representation of Vector Electromagnetic Beams, Phys. Rev. A {\bf 22}, 1159-1164 (1980).

\bibitem{Davi-P}
L. W. Davis and G. Patsakos, Comment on ``Representation of Vector Electromagnetic Beams'', Phys. Rev. A {\bf 26}, 3702-3703 (1982).


\bibitem{Akhi-B}
A. I. Akhiezer and V. B. Berestetskii, {\it Quantum Electrodynamics} (Interscience, New York, 1965).

\bibitem{Cohe-DG}
C. Cohen-Tannoudji, J. Dupont-Roc, and G. Grynberg, {\it Photons and Atoms} (John Wiley \&
Sons, New York, 1989).

\bibitem{Bomz-BK}
Z. Bomzon, G. Biener, V. Kleiner, and E. Hasman, Radially and azimuthally polarized beams generated by space-variant dielectric subwavelength gratings, Opt. Lett. {\bf 27}, 285-287 (2002).

\bibitem{Tung}
W.-K. Tung, {\it Group Theory in Physics} (World Scientific, Singapore, 1985).

\bibitem{Golu-L}
G. H. Golub and C. F. Van Loan, {\it Matrix Computations}, 3rd ed. (Johns Hopkins, Baltimore, 1996).

\bibitem{Mess}
A. Messiah, {\it Quantum Mechanics}, Vol. II (North Holland Publishing Company, Amsterdam, 1962).

\bibitem{Merz}
E. Merzbacher, {\it Quantum Mechanics}, 3rd ed. (John Wiley \& Sons, New York, 1998).


\bibitem{Lawr-P}
L. W. Davis and G. Patsakos, TM and TE electromagnetic beams in free space, Opt. Lett. {\bf 6}, 22-23 (1981).

\bibitem{Li15}
C.-F. Li, On photon angular momentum: transversality condition, Berry degree of freedom, and non-commutativity of photon position. Preprint at https://arxiv.org/pdf/1501.06672 (2015).

\end{thebibliography}

\end{document}